\documentclass[twocolumn,aps,floatfix,showpacs,superscriptaddress,nofootinbib]{revtex4-1}
\usepackage{bm,amsmath,amssymb,graphicx}

\begin{document}

\title{Effective ergodicity breaking in an exclusion process with varying system length}

\author{Christoph Schultens}
\affiliation{Institut f\"{u}r Theoretische Physik, Universit\"{a}t zu K\"{o}ln, 50937 K\"{o}ln, Germany}
\author{Andreas Schadschneider}
\affiliation{Institut f\"{u}r Theoretische Physik, Universit\"{a}t zu K\"{o}ln, 50937 K\"{o}ln, Germany}
\author{Chikashi Arita}
\affiliation{Theoretische Physik, Universit\"at des Saarlandes,  66041 Saarbr\"ucken, Germany}


\begin{abstract}
Stochastic processes of interacting particles 
  in systems   with varying length
  are relevant e.g. for several biological applications.  We try to
  explore what kind of new physical effects one can expect in such
  systems.  As an example, we extend the exclusive queueing process
   that can be viewed as a one-dimensional exclusion process with
  varying length, by introducing Langmuir kinetics.  This process can
  be interpreted as an effective model for a queue that interacts with
  other queues by allowing incoming and leaving of customers in the
  bulk.  We find surprising indications for breaking of ergodicity in
  a certain parameter regime, where the asymptotic growth behavior
  depends on the initial length.  We show that a random walk with
  site-dependent hopping probabilities exhibits qualitatively the same
  behavior.  
\end{abstract}

\maketitle 

\section{Introduction}  
The exclusive queueing process (EQP) 
\cite{bib:A,bib:YTJN,bib:AY,bib:ASPED14,bib:M3AS} is a
queueing model that takes into account the spatial structure of the
queue.  In standard queueing theory, which is a well-established
approach of practical relevance \cite{bib:E,bib:K,bib:Sa}, the system
length and the number of particles are identical, so that the density
along the queue is always 1.  In the EQP, particles interact with each
other through an exclusion principle, i.e.  they can move forward only
when the  
  target  site is empty. Thus the density is not a constant.
The EQP is equivalent to the totally asymmetric simple exclusion
process (TASEP) \cite{bib:L,bib:Schuetz,bib:CSS,bib:SCN} of varying
length.  One end of the chain is fixed and corresponds to the server
to which particles move.  Arriving particles always join the queue at
the site just behind the opposite end of the queue.  In earlier works
\cite{bib:AY,bib:AS1} the phase diagram of the EQP with parallel
update scheme has been determined exactly. It shows two main phases
that correspond to queues with converging and diverging lengths.
These phases can further be divided into several subphases, see
\cite{bib:AS1,bib:AS2,bib:AS3} for details.

In the previous work \cite{bib:AS4}, unusual critical behavior of the
EQP has been observed. This indicates that dynamical systems of
fluctuating length might show surprising properties.  Due to the
relevance of stochastic processes with varying system length, in
particular for applications to biology
\cite{bib:SEPR,bib:SE,bib:ES,bib:DMP,bib:JEK,bib:MRF,bib:SS,bib:M,bib:dGF},
it is worthwhile to explore them in more detail.  Extending our
previous works, here we introduce an EQP with Langmuir kinetics
(EQP-LK) that allows creation and annihilation of particles anywhere
in the bulk of the queue, not only at the ends.  This can be
interpreted as an effective model for interacting queues.   For
  customer queues, customers might change to (from) the bulk of the
  queue from (to) other queues.  Looking at a specific queue this has
  the same effects as Langmuir kinetics characterized by a detachment
  probability $\omega_D$ and an attachment probability $\omega_A$ (if
  every vacancy in the queue can be occupied by an arriving
  customer).

In this work, we shall characterize the EQP-LK via the system length
$L_t$ at time $t$.  We begin by introducing a naive test to determine
whether the system length diverges or converges to a stationary length
by averaging simulation samples with initial condition $L_{t=0}=0$.
Next we shall examine the time evolution of the average system length
$\langle L_t\rangle$.  Surprisingly its behavior depends on the
initial system length $L_0=L_{t=0}$, indicating the breaking of
ergodicity.  Furthermore, we shall see that individual simulation
samples can exhibit different behaviors.  We shall qualitatively
explain these unexpected phenomena by constructing a random walk model
which captures the essential features of the length dynamics.  We
expect that these new insights could be of relevance also for the
interpretation of experimental studies on systems of varying length,
especially in biology.


\section{Model} 

\begin{figure} 
\begin{center}
 \includegraphics[width=\columnwidth]{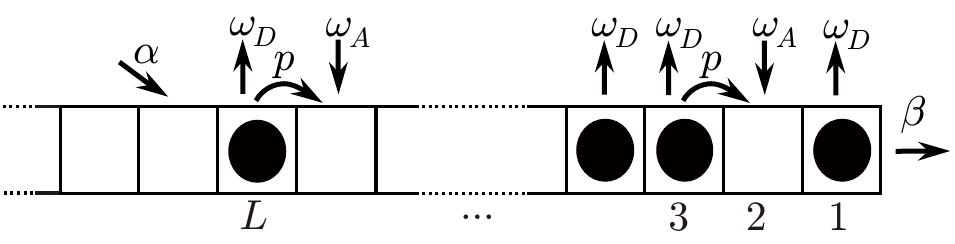}
\end{center}
\caption{Transition probabilities of 
  the exclusive queueing process with Langmuir kinetics (EQP-LK).}
\label{fig:EQPLK} 
\end{figure}

The model that we study is a combination of the EQP with parallel
update \cite{bib:AY} and Langmuir kinetics (EQP-LK), see
Fig.~\ref{fig:EQPLK}. It is defined on a semi-infinite one-dimensional
lattice, where the sites are numbered from right to left. Each site
$j\in \mathbb N$ is either occupied by a particle ($\tau_j=1$) or
empty ($\tau_j=0$).  Our model's state space is infinite but
countable, consisting of configurations $\tau_L\cdots \tau_1$ and a
state $\emptyset$ where there is no particle. The system length $L$ is
defined by the leftmost occupied site or $L=0$ for the state
$\emptyset$.  The system length is, in general, different from the
number of particles, in contrast to classical queueing models.
Particles move forward (rightward in Fig.~\ref{fig:EQPLK}) with
probability $p$ in each time step only if the preceding site is
unoccupied.  A new particle enters at the end of the queue (i.e.
$j=L+1$) with probability $\alpha$.  When there is no particle in the
system, a new particle enters directly at site $j=1$.  Particles in
the bulk are detached with probability $\omega_D$, and for each empty
site $j(\le L)$ a particle is attached with probability $\omega_A$.
As in the TASEP with Langmuir kinetics (TASEP-LK)
\cite{bib:PFF,bib:Popkov,bib:EJS,bib:DG}, the attachment and detachment 
probabilities are scaled with the system length as $\omega_A = \Omega_A/L$ 
and $\omega_D = \Omega_D/L$.
 The total probabilites $\Omega_A$ and $\Omega_D$ are kept constant.
Then inflow and outflow caused by the Langmuir kinetics are 
given by $  (1-\rho) \Omega_A $  and  $ \rho \Omega_D $,
where $\rho$ is the global density of the system. It is of the
same order of magnitude as the boundary currents so that the
dynamics of the system is determined by a competition between bulk and
boundary dynamics.   
In contrast to the TASEP-LK, the system length $L$ of the EQP-LK varies, 
and thus the probabilities $\omega_A, \omega_D$ depend on the current state
 and thus has to be determined from the fixed values of 
$\Omega_A, \Omega_D$ as $\omega_A = \Omega_A/L$ and $\omega_D = \Omega_D/L$. 
In each time step, first the configuration is
updated according to the rule of the EQP with parallel update, and
then the Langmuir kinetics is applied. This defines the EQP-LK with
parameters $(p,\alpha,\beta,\Omega_A,\Omega_D)$,
which is generically an irreducible and aperiodic discrete-time Markov
process.  We denote by $L_t$ the system length at time $t$ for a
realization (a simulation run) of the stochastic process and by
$\langle L_t\rangle$ its average over different realizations.


\section{Phase diagram} 

First we revisit the EQP with parallel update, corresponding to the
special case of the EQP-LK with $\Omega_A=\Omega_D=0 $.  The parameter
space is divided into two regimes by the ``critical line''
\begin{equation}
\alpha_c  =\begin{cases}
\beta(p-\beta)/(p-\beta^2) & (\text{for }\beta\le 1-\sqrt{1-p} ),  \\ 
 (1-\sqrt{1-p})/2 &  (\text{for }\beta>1-\sqrt{1-p} ) .
\end{cases}
\label{eq:alphac}
\end{equation}
For $\alpha<\alpha_c$ (``convergent phase''), the average length
$\langle L_t \rangle $ converges to  a stationary value.
On the other hand, for $\alpha>\alpha_c$ (``divergent phase''),
$\langle L_t \rangle $ diverges.  These properties were shown
rigorously by constructing an exact stationary distribution
\cite{bib:AY}.  Furthermore, in the divergent phase, the asymptotic
behavior of $\langle L_t \rangle $ was shown to be linear in time
$\langle L_t \rangle = Vt $ with an explicit form for the velocity
$V$, by simulations for general $p$ \cite{bib:AS3} and by using an
exact time-dependent solution for $p=1$ \cite{bib:AS2}.

\begin{figure}
   \includegraphics[width=\columnwidth]{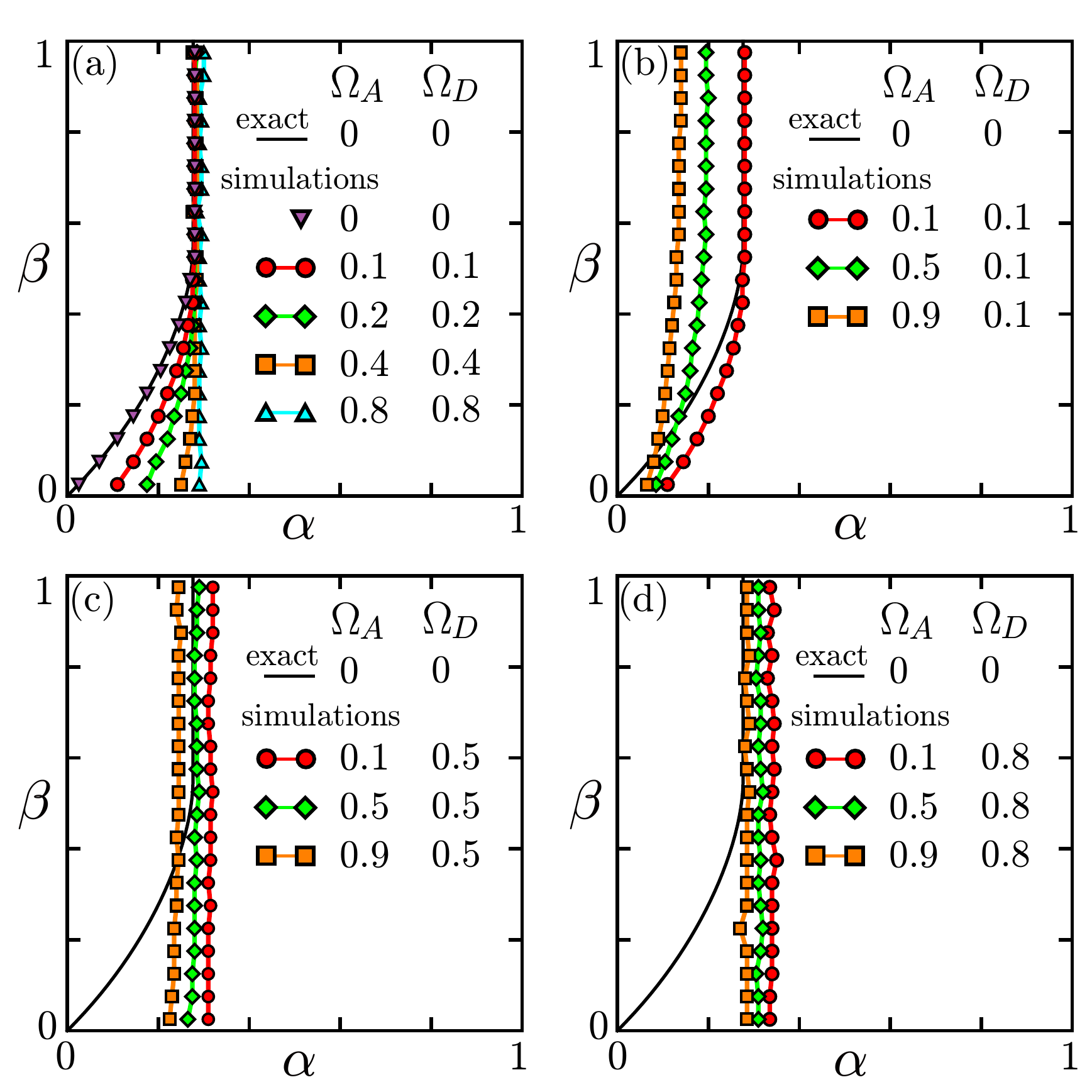}
\vspace{-5mm}
\caption{Phase diagrams for
  various values of $\Omega_D, \Omega_A$ and $p=0.8$, determined by
  the test \eqref{eq:test}.  For comparison, the
  exact critical line \eqref{eq:alphac} for the EQP case
  $\Omega_D=\Omega_A=0$ is shown, which is recovered by the test.}
\label{fig:phase0} 
\end{figure}

Now we introduce a test to distinguish between diverging and
converging system lengths by simulations.  Starting from $L_{t=0}=0$,
the quantity
\begin{align}
\label{eq:R=}
R_T =\sum_{3T/4<t\le T} \langle L_t  \rangle \big/
    \sum_{T/2<t\le 3T/4}  \langle L_t  \rangle  
\end{align} 
approaches  $R_\text{conv}=1$ as $T\to\infty$ in the convergent
phase.  On the other hand, if we assume $\langle L_t \rangle$ diverges
linearly in time, the quantity $R_T$ approaches
  $R_\text{div}=7/5$.  In computer simulations only finite $T$ can
be studied. Here we set $T= 2\cdot 10^4$.  Averages are calculated
with a finite number $10^3$ of samples.  The phase transition is
identified as the point where $R$ becomes bigger than the average 
\begin{equation}
\label{eq:test}
R_T >   (R_\text{conv} + R_\text{div} ) /2 = 6/5 
\end{equation}
when $\alpha$ is increased with the other parameters fixed\footnote{The value 6/5 does not reflect the true value of  $\lim_{T\to\infty}R_T$ 
  on the phase transition line (see e.g. \cite{bib:AS4}).
 However, it provides a convenient criterion that allows to distinguish 
 (linear) divergence from convergence in the simulations.  }.
In Fig.~\ref{fig:phase0}(a), we observe that for the EQP the critical
line obtained by this test agrees with the exact line.

Let us apply the test to the EQP-LK, again starting from the initial
state $\emptyset$.
The phase boundary between the two phases depends on the Langmuir
probabilities $\Omega_A$ and $\Omega_D$, see Fig.~\ref{fig:phase0}.
When the ratio $\Omega_A/\Omega_D $ is 1, 
  we observe that the phase boundary becomes 
 simply a straight segment $\alpha=(1-\sqrt{1-p})/2$
as $\Omega_A=\Omega_D\to 1$, see Fig.~\ref{fig:phase0} (a). 
 For fixed
$\Omega_D$, the natural observation is that the convergent phase is
enlarged for increasing values of $\Omega_A$, see
Fig.~\ref{fig:phase0} (b), (c) and (d).  Again the critical line
becomes a straight segment (which is independent of $\alpha$) as
$\Omega_D$ increases.


\section{Dependence on initial conditions}

\begin{figure}
  \includegraphics[width=\columnwidth]{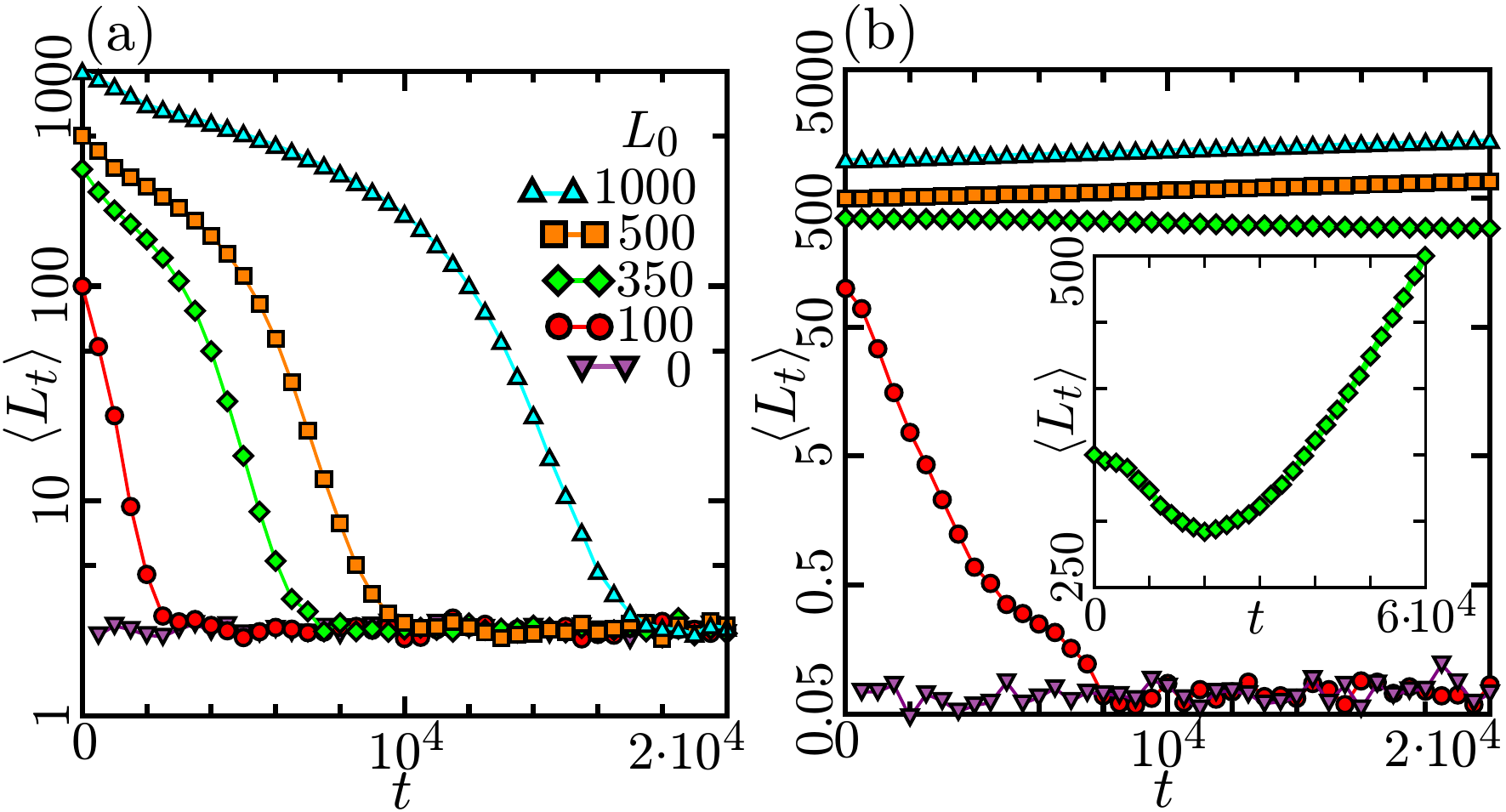}
\caption{
  Behavior of $\langle L_t\rangle$ starting from various initial lengths,
 for the parameters given in Eqn.~\eqref{eq:specific-parameters-conv}(a)
  and  Eqn.~\eqref{eq:specific-parameters}(b). 
   We have  set the initial density as $\Omega_A/(\Omega_D+\Omega_A)$, and
  averaged over $10^3$ samples. In (a), all the average lengths ($
  L_0=0,100,350,500,1000 $) converge to a stationary value. In (b),
  we observe  not all the average lengths converge.  }
\label{fig:ergo-nonergo} 
\end{figure}
The EQP-LK with generic values of parameters is an irreducible and
aperiodic Markov process on a countable state space. The general
theory of Markov processes \cite{bib:Schinazi} tells us that the
convergence of $\langle L_t \rangle $ is independent of the initial
state. We now check this property by simulations.  For example, the
case
\begin{align}
  \label{eq:specific-parameters-conv}
  (p,\alpha,\beta,\Omega_A,\Omega_D)=(0.8,0.2,0.2,0.2,0.35)
\end{align} 
is determined to be in the convergent phase by the test (\ref{eq:test}).
In fact, the average lengths $\langle L_t \rangle $ over $10^3$
simulation samples with other initial lengths $L_0$ converge to
the same stationary value, see Fig.~\ref{fig:ergo-nonergo}(a).
Surprisingly, however, this is not always true.  For example, for 
\begin{align}\label{eq:specific-parameters}
(p,\alpha,\beta,\Omega_A,\Omega_D)=(0.8,0.3,0.2,0.1,0.9),
\end{align}
the average length $\langle L_t \rangle$ with $L_0=0$ converges but
the average lengths with large initial lengths (e.g. $L_0=500,1000$)
do not converge, see Fig.~\ref{fig:ergo-nonergo}(b). Furthermore,
starting from an intermediate length (e.g.\ $L_0=350$), the average
length $\langle L_t \rangle$ exhibits a non-monotonic behavior. In the
inset of Fig.~\ref{fig:ergo-nonergo}~(b), $ \langle L_t \rangle$
decreases until $t\sim 2\cdot 10^4$ and then increases.  In the rest
of this work, we shall try to understand this unexpected phenomenon
with the specific set of parameter values
\eqref{eq:specific-parameters}.  We emphasize that this
unexpected behavior is observed generically in a larger parameter
regime \cite{BA1,BA2}.

\begin{figure}
\begin{center}
   \includegraphics[width=\columnwidth]{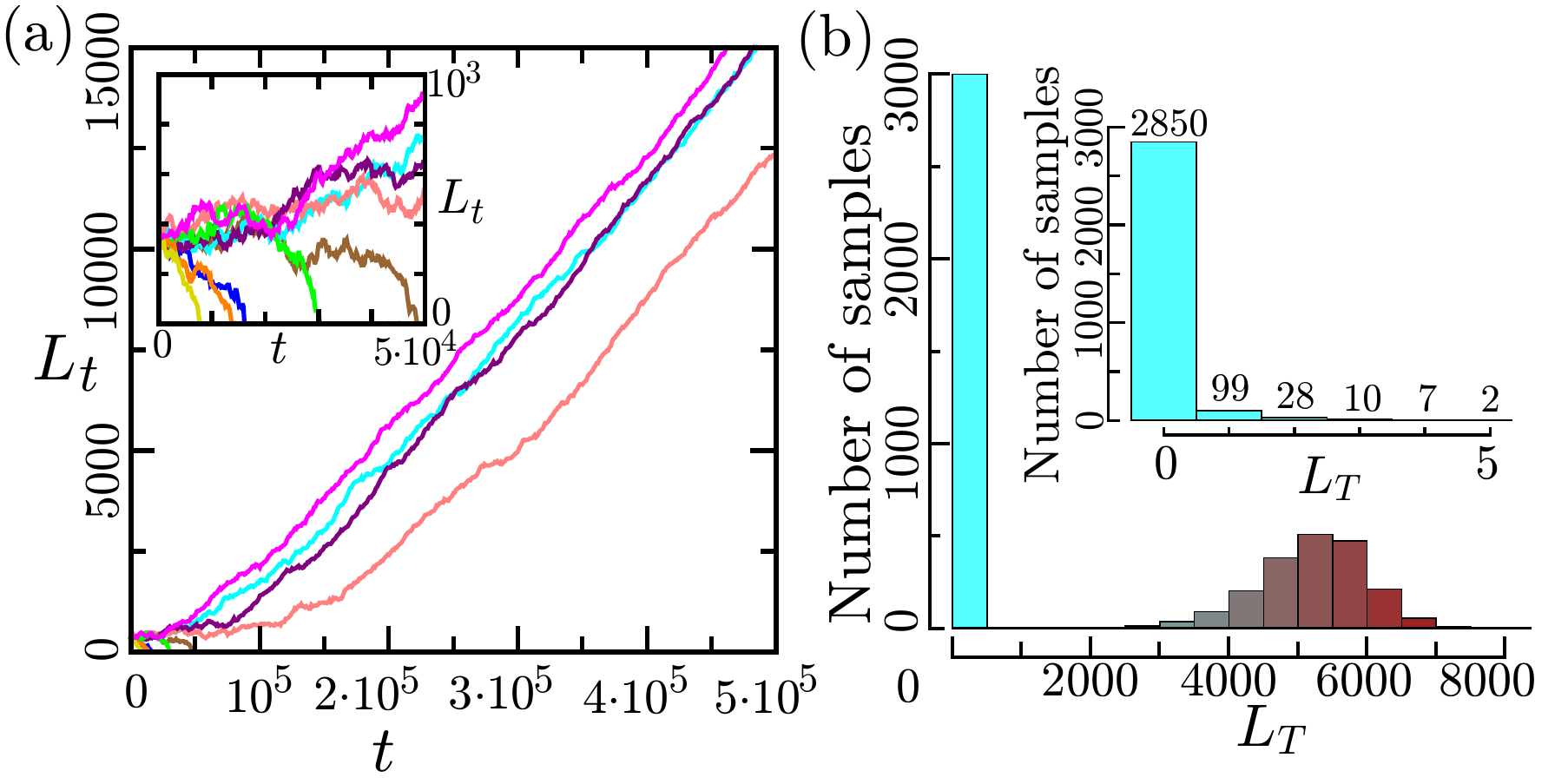}
\end{center}
\caption{\label{fig:L0350}
  (a) Behaviors of $L_t$ for 9 individual samples and (b) the length
  distribution of $5\cdot 10^3$ samples at time $t=T=2\cdot 10^5$.
  The parameters were set as Eqn.~\eqref{eq:specific-parameters}, and
  every sample started from the length $L_0=350$ and the density $
  \Omega_A/(\Omega_A+\Omega_D)=0.1 $.  More precisely, the bars in (b)
  represent the number of samples that satisfy $L_T \in [ 0,500 ), [
  500,1000 ), \ldots$.  The inset shows more detailed distribution
  near $L_T=0$, i.e. the bars represent the numbers of samples such
  that $ L_T=0,1,\ldots,5$.}
\end{figure}

First, let us investigate the behavior of {\it individual}
samples.  Comparing the insets of Fig.~\ref{fig:ergo-nonergo}(b) and
Fig.~\ref{fig:L0350}(a), we notice that the individual behaviors do
not perfectly mimic the behavior of the average. In other words, the
average does not represent the {\it typical} behavior of individual
samples.  Therefore the dynamics of the system cannot be properly
understood by just looking at averages.  
 For example, on the ``coexistence line'' of the TASEP with open
boundaries, the average density profile is linear.  However, this does
not imply that a shock is moving, which can only be observed for
individual samples \cite{bib:KSKS}.

In Fig.~\ref{fig:L0350}(a), we observe that 5 of 9 samples hit $L_t=
0$ within $t<5 \cdot 10^4$. We call them {\it converging samples}. The
other 4 samples increase almost linearly in time, even after $ t\sim
10^5$ ({\it diverging samples}).  Let us look at statistics of $5\cdot
10^3$ samples with the same parameter setting and the same initial
length [Fig.~\ref{fig:L0350}(b)].  Apparently there are two peaks at
$L_T\in [0,500)$ and $L_T\in [5000,5500)$, corresponding to converging
and diverging samples, respectively.  Because of the strong effect of
the detachment $\Omega_D=0.9$, it is difficult to escape from $ L_t=0
$ after reaching $ L_t=0 $, see the inset of in
Fig.~\ref{fig:L0350}(b).  The non-monotonicity observed for the
average $\langle L_t\rangle$ is due to the dominance of the
contributions from the diverging samples whereas the contribution of
the converging can be neglected once they have reached $L_t=0$.  We
note that $L_t=0$ is not an absorbing state, and our model exhibits no
absorbing transition which was studied in a symmetric exclusion
process with varying length \cite{bib:BH}.

\begin{figure}[tb] 
\begin{center}
 \includegraphics[width=\columnwidth]{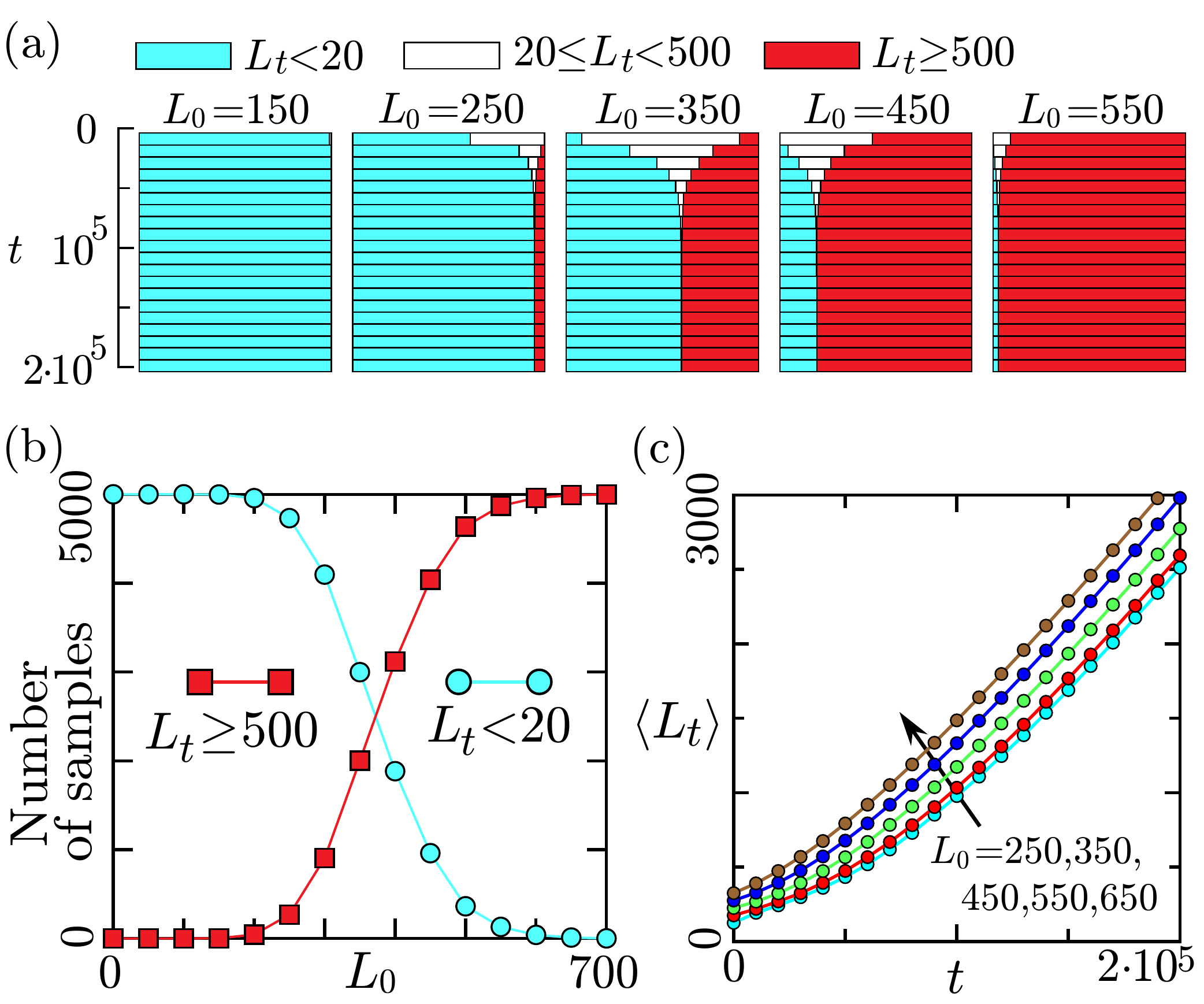}
\end{center}
\caption{
  (a,b) System length $L_t$ distributions of $5\cdot 10^3$ simulation
  samples, and (b) behavior of $L_t $ averaged over diverging samples,
  with parameters \eqref{eq:specific-parameters}.  The samples started
  from various initial lengths at density 
  $\Omega_A/(\Omega_A+\Omega_D)=0.1$.
  In (a) the samples are divided
  into three groups according to $L_t \in [0,20), [20,500)$ or
  $[500,\infty)$ at time $t=10^4,2\cdot 10^4,3\cdot 10^4,\dots, 2\cdot
  10^5$. The lengths of bands represent the ratio of the number of
  samples classified into each group.  The plots of the number of
  samples at $t=2\cdot 10^5 $ are given in (b).  The plots in (c) are
  given by averaging over samples that have never hit $L_t=0$ within
  $t\le 2\cdot 10^5$.  }
\label{fig:variousL0}
\end{figure}

Distributions of simulation samples for various initial lengths is
provided in Fig.~\ref{fig:variousL0}(a,b).  We observe that the
samples starting from a short queue tend to remain short. More samples
starting from a long queue tend to grow.  The growth of each of them
is almost proportional to time $t$ as well as their average, see
Fig.~\ref{fig:L0350}(a) and Fig.~\ref{fig:variousL0}(c).

\section{First passage time} 

\begin{figure} 
\begin{center}
  \includegraphics[width=\columnwidth]{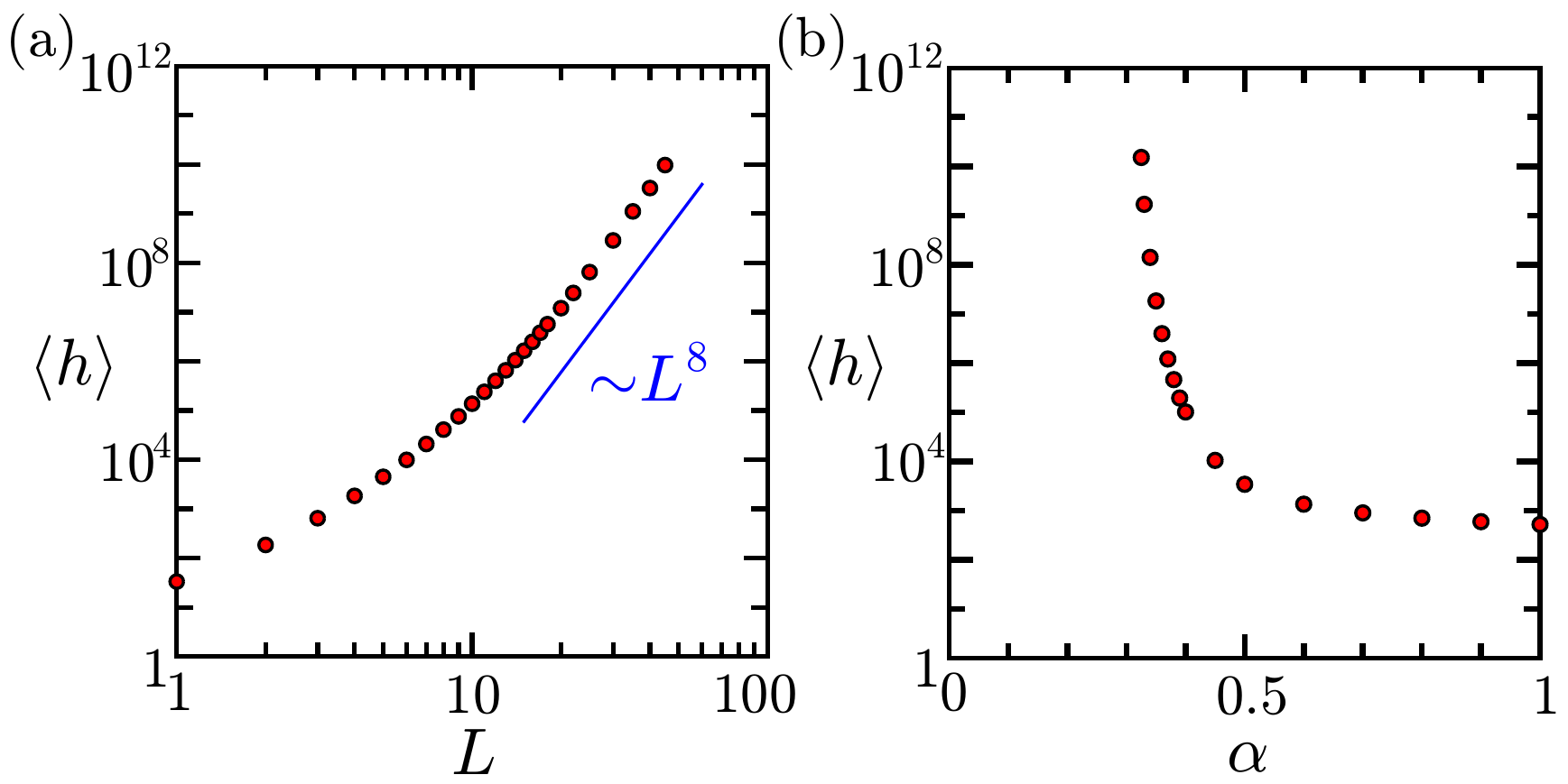}
\end{center}
\caption{(a)  Mean first passage time $\langle h\rangle $ (when 
  samples visit $ L_h =L$ for the first time) vs $L$.  The parameters
  were chosen as in Eqn.~\eqref{eq:specific-parameters}   
  with initial length $L_0=0$. Averages were taken over $10^4$
  (for $L\le 30$) or $500$ (for $L>30$) simulation samples.  For
  comparison the line const.$\times L^8$ is shown.  (b) Mean first
  passage time $\langle h\rangle $ when samples visit $ L_h =500$ for
  the first time vs $\alpha$.  The other parameters were chosen as
  $(p, \beta,\Omega_A, \Omega_D) = (0.8,0.2,0.1, 0.9)$.   Averages
    were taken  over $10^4$ (for $\alpha\ge 0.35$), $10^2$ (for
  $\alpha= 0.33,0.34$) or $10$ (for $\alpha=0.325$) simulation samples.
   }
\label{fig:hit}
\end{figure}

Let us consider the first passage time \cite{bib:R}, i.e. the first
time $h$ when a sample hits the length $L_h=L$, starting from $L_0=0$.
We observe that the average first passage time becomes very large as
$L$ increases [Fig.~\ref{fig:hit} (a)], and it seems to increase
faster than power law.  Thus it is impossible to reach e.g.  $L_t =
500 $ in our computer simulations, even though the probability of {\it
  ever} hitting the length $L_t = 500$ is 1.  In this sense, the
ergodicity of the EQP-LK is {\it effectively} broken\footnote{In
  \cite{bib:RPS}, the authors introduced Langmuir-kinetics like
  attachment and detachment into the TASEP on a finite chain.  This
  model's ergodicity is also broken   in the sense that each simulation sample
  switches between two different density profiles with a long
  lifetime.  Similarly, a mass-transport model with continuous state
  variables has been investigated in \cite{bib:Zielen}, where each
  sample switches between a high-flow and a low-flow state.}.  The
difficulty of reaching $L_t = 500$ is also implied by
Fig.~\ref{fig:hit} (b), where the average first passage time becomes
extremely large as $\alpha $ decreases.


\section{Random walk model}
\begin{figure}    
\begin{center}
 \includegraphics[width=0.7\columnwidth]{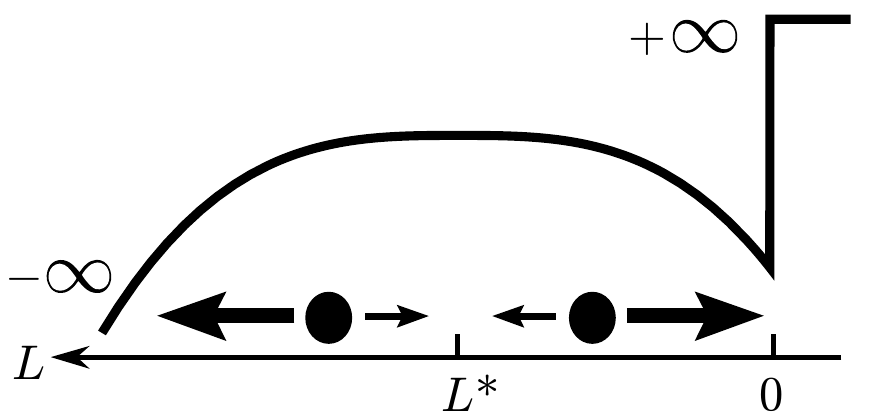} 
\end{center}
 \caption{An inhomogeneous random walk model with one reflecting 
   boundary (at $L=0$).  When the walker is on a position $L<L^*$
   ($L>L^*$), it prefers to go to $L=0$ (resp. $L=+\infty$).
   }\label{fig:potential}\end{figure}

 The mechanism underlying the effective ergodicity breaking can be
qualitatively understood in terms of a random walk model.  The
position of the walker corresponds to the system length $L$ and the
random walk has one reflecting end corresponding to $L=0$. We denote
the hopping probabilities by $q_L $ for $L\to L+1 $ and by $r_L $ for
$L+1\to L$.  The behavior of the length of the EQP-LK can be
qualitatively modeled by hopping probabilities that satisfy
\begin{align}
    \frac{q_L}{r_L}
    \begin{cases} 
       < 1 &  (L<L^*), \\ 
       = 1 & (L = L^*), \\ 
        > 1 & (L> L^*)
    \end{cases}
\end{align}
with some $L^*$.
In other words, the potential \cite{bib:RPS} $U(L) = \sum_{j=0}^{L-1}
\ln \frac{r_j}{q_j}$   takes a maximum at $L=L^*$, 
and $\lim_{L\to\infty}U(L)=-\infty $,
see Fig.~\ref{fig:potential}.
One of the simplest examples is the case where
$ q_L=1-r_{L-1} = \frac{1}{2\pi} \arctan (\frac{ L-L^* }{c}) +\frac{1}{2}$ 
with some $c$.

The walker tends to move towards $L=0$ when it is on a position
$L<L^*$. Oppositely a walker on $L>L^*$ tends to move towards
$L=+\infty$.  This inhomogeneous bias corresponds to the length
dependence of the Langmuir probability; when the system length is
short, a newly entering particle   at the left end, which would increase the length, 
can easily be removed.  On the other
hand, in a long queue this effect is weak  since  $\Omega_D/L
\approx 0$.  Therefore the initial condition and initial behavior are
very important. Once the walker reaches $L=0$, it cannot easily escape.
  It then fluctuates near $L=0$, but this
is not a true stationary state.  We remark that one can easily prove
that there is no stationary distribution, and the walker's position
should diverge in the long-time limit.  Thus the random walk model
exhibits qualitatively the same behavior as observed in the EQP-LK
(see Figs.~\ref{fig:L0350} and \ref{fig:variousL0}).

\section{Discussion}
We have analyzed a queueing model with excluded-volume effect and
Langmuir kinetics by simulations.  Due to the varying length of the
system, the Langmuir probabilities depend on the current state in each
simulation run, which has a significant influence on the dynamics,
e.g.  a strong dependence on the initial condition and effective
ergodicity breaking.  There is a phase where long queues ($L>L^*$)
prefer to grow, whereas short queues ($L<L^*$) prefer to remain short,
although full identification of such regime has not yet been completed
\cite{BA1,BA2}.

The EQP-LK shows that stochastic systems on fluctuating geometries can
exhibit surprising behavior. We believe that beyond the theoretical
interest our findings could be relevant for the interpretation of
experimental (where the number of samples is necessarily finite)
results as well, especially in biological systems.

\vspace{0.5cm}

\noindent{\bf Acknowledgements:}\\
We are grateful to Christian Borghardt for his
contributions in an early stage of this work.
We also thank Martin R. Evans and Joachim
Krug for useful discussions. This work was partially supported by
Deutsche Forschungsgemeinschaft (DFG) under grant ``Scha 636/8-1''.

\end{document}